# The small $x$ behaviour of $g_1$

S. D. Bass[1] and P. V. Landshoff[2]

[1]*HEP Group, Cavendish Laboratory,*
*University of Cambridge, Madingley Road, Cambridge CB3 0HE, England*

[2]*Department of Applied Mathematics and Theoretical Physics, University of Cambridge,*
*Silver Street, Cambridge CB3 9EW, England*

## Abstract

We discuss the small $x$ behaviour of the spin dependent structure function $g_1$. We find a contribution from the exchange of two non-perturbative gluons which behaves as $\sim (2\ln\frac{1}{x} - 1)$.

In recent years there has been much interest in polarised deep inelastic scattering as a result of the European Muon Collaboration (EMC) measurement of the polarised proton structure function $g_1^P(x, Q^2)$ [1]. The naive parton model interpretation of the EMC data is that the quarks contribute a small fraction of the proton's spin.

The EMC experiment, which followed earlier polarised scattering experiments at SLAC [2], has inspired a new experimental programme. During 1993-1994 we have been presented with new data from the Spin Muon Collaboration (SMC) at CERN [3,4] and the E-142 experiment at SLAC [5]. New experiments are planned in the near future [6]. One of the key ingredients in using deep inelastic scattering experiments to test spin sum-rules (and hence learn about the spin structure of the nucleon) is to extrapolate the $g_1$ structure function data to $x = 0$. This extrapolation introduces a theoretical error which must be included in the analysis.

There are two sum-rules which are important in polarised deep inelastic scattering. The first is the Bjorken sum-rule [7] for the iso-triplet part of $g_1$

$$\int_0^1 dx \left(g_1^p - g_1^n\right)(x, Q^2) = \frac{1}{6} g_A^3 \, C_3^q(Q^2) \tag{1}$$

Here $g_A^3$ is the iso-triplet axial charge and $C_3^q$ is the non-singlet perturbative Wilson coefficient, which has been evaluated to $O(\alpha_s^3)$ precision [8]. The Bjorken sum-rule was derived using current algebra before the advent of QCD and is a test of isospin. It is not expected to fail.

The Ellis-Jaffe sum-rule [9] is a test of Zweig's rule in the flavour singlet channel. If we assume that strange (and heavy) quarks do not play a significant role in polarised deep inelastic scattering, then we can derive sum-rules for each of the proton ($\int_0^1 dx g_1^p$) and neutron ($\int_0^1 dx g_1^n$) targets. The first moment of the flavour singlet part of $g_1$ is

$$\int_0^1 dx \, g_1(x, Q^2)|_S = \frac{1}{3}\sqrt{\frac{2}{3}} \, g_A^0 \, C_0^q(Q^2) \tag{2}$$

Here $g_A^0$ is the flavour singlet axial charge and $C_0^q$ is the singlet coefficient, which has been calculated to $O(\alpha_s^2)$ precision [10]. In the naive parton model $\Sigma = \sqrt{6} g_A^0$ is interpreted as the quark spin content

of the nucleon. The most recent determination of this quantity (which includes the SMC proton data [4]) is

$$\Sigma = 0.30 \pm 0.07(stat.) \pm 0.10(syst.) \qquad (3)$$

This number is two standard deviations below the naive quark model expectation ($\Sigma = 0.68$), which is based on the assumption that Zweig's rule is exact. In QCD one finds that the spin of the quarks is screened by the gauge symmetry of the background colour field in the nucleon (the physics of the axial anomaly). This result means that $g_1$ does not have a simple explanation in terms of explicit quark and gluon spin degrees of freedom (for reviews see [11-13]).

The quantity which is measured in polarised experiments is the spin asymmetry $A_1$. In order to test spin sum-rules for the first moment of $g_1$ we first need to extract the structure function from the $A_1$ data and then extrapolate $g_1$ to $x = 0$ using Regge theory, which provides a good description of the small $x$ behaviour of the unpolarised structure function in the NMC kinematic range.

The experiments which have been carried out to date use a longitudinally polarised beam and target [14]. The spin structure function is related to the spin asymmetry $A_1$ via the equation:

$$g_1(x, Q^2) = \frac{A_1(x, Q^2) F_2(x, Q^2)}{2x(1 + R(x, Q^2))} \qquad (4)$$

In order to extract $g_1$ from $A_1$, the experimenters use the NMC parametrisation of $F_2$ [15] and the SLAC parametrisation of $R(x, Q^2)$, which was determined by Whitlow et al. [16]. This parametrisation of $R(x, Q^2)$ was obtained from a fit to data on the longitudinal structure function $F_L$ in the region $x > 0.1$. It includes the effects of leading order QCD evolution and twist four terms. At the present time, the smallest $x$ data point for $g_1^p$ (which comes from SMC) has been measured at $<x> = 0.005$ and $<Q^2> = 1.3$ GeV$^2$ [4]. This measurement is obtained from $A_1(x, Q^2)$ via an extrapolation of $R(x, Q^2)$ from $x = 0.1$ down to $x = 0.005$. It seems reasonable to us that one should question the size of the error that results from using this extrapolation of $R(x, Q^2)$ to such small $x$ and $Q^2$. One should also bear in mind possible shadowing corrections when one extracts $g_1^n$ from measurements on either a deuteron target (SMC) or a $^3He$ target (E-142). Shadowing in $F_2$ in the deuteron is typically 3-6% at $<x> = 0.005$ [17] and may be a factor of two greater in $g_1$ [18].

Given that we have extracted the structure function $g_1$ from $A_1$ how should we extrapolate it to $x = 0$ ? It is well known [19, 20] that the iso-triplet piece of $g_1$, which appears in the Bjorken sum-rule, behaves as

$$g_1^{(3)} \sim x^{-\alpha_{a_1}}, \quad x \to 0 \qquad (5)$$

where $\alpha_{a_1}$ is the intercept of the $a_1$ Regge trajectory. Heimann quotes this intercept as $\alpha_{a_1} = -0.14$ [19], which is within the phenomenological range $-0.5 \leq \alpha_{a_1} \leq 0$ discussed by Ellis and Karliner [21]. If one assumes that the slope of the $a_1$ trajectory is the same as the slope of the $\rho, \omega$ trajectory, then one finds that $\alpha_{a_1} \approx -0.4$. As a model-independent statement, one can say that the iso-triplet part of $g_1$ converges as $x \to 0$.

Usually $g_1$ is extrapolated to small $x$ as $g_1 \sim constant$, which is consistent with assuming that $g_1$ is almost pure iso-triplet at small $x$ if we take Heimann's value of $\alpha_{a_1}$. On the other hand, the latest $g_1^p$ data from SMC show evidence of a possible rise in $g_1$ in the smallest $x$ bins [4]. In the rest of this paper we discuss the small $x$ behaviour of the flavour singlet part of $g_1$ and report on a gluonic exchange contribution to spin-dependent processes which has hitherto been missed in the Regge spin literature.

In the flavour singlet channel we have to consider gluonic exchanges. The small $x$ piece of the unpolarised structure function $F_2$ is governed by pomeron exchange. The NMC data at small $x$ (in the range $0.005 < x < 0.1$) are described by the physics of the non-perturbative pomeron [22]. Strictly speaking, the non-perturbative pomeron is relevant to deep inelastic scattering in that it describes the total cross section for a real or almost-real high-energy photon scattering from the nucleon. At very small $x$ (eg. at HERA) the structure function $F_2$ rises faster than the prediction of simple non-perturbative pomeron exchange [23]. This tells us that off-shell (finite $Q^2$) effects are important in the range of the HERA kinematics and may be evidence of a perturbative pomeron [24].

At the present time, we have data on the spin dependent structure function $g_1$ for $x > 0.005$; that is, in the same $x$ range as the NMC unpolarised structure function data. Whilst the usual pomeron (with

Regge intercept +1) does not contribute to $g_1$ [19,20], it seems reasonable that the *physics* which leads to pomeron exchange may also be important in understanding the small $x$ behaviour of $g_1$. Indeed, we find that nonperturbative two-gluon exchange, which seems to be an important component of the nonperturbative pomeron [26-29], contributes to $g_1$ a small $x$ behaviour that is roughly constant.

The perturbative parton model suggests that there is more to the small $x$ behaviour of $g_1$ than the iso-triplet $a_1$ exchange in equ.(5). In the parton model the spin dependent gluon distribution $\Delta g(x, Q^2)$ contributes to the flavour singlet part of $g_1$ as $\Delta g \otimes C^g$, where $C^g$ is the gluonic Wilson coefficient. This gluonic contribution is clearly dependent on the shape of $\Delta g(x, Q^2)$ but can easily lead to a contribution which diverges as $x \to 0$ [25]. At the present time, there is no experimental measurement of $\Delta g(x, Q^2)$ or even the sign of $\Delta g = \int_0^1 dx \Delta g(x, Q^2)$.

The simple and successful four parameter model of non-perturbative pomeron exchange involves the exchange of two non-perturbative gluons [26-29]. This is shown for deep inelastic scattering in Fig. 1. Here $p$ is the nucleon target momentum, $q$ is the photon momentum and we use $x_{Bj}$ to denote the Bjorken variable. (The crossed-quark-box graph gives only non-leading terms as $\nu = p.q \to \infty$.) We use the Sudakov variables:

$$k = xp + yq + k_T$$
$$l = \xi p + \overline{\eta}\frac{q}{2\nu} + l_T \qquad (6)$$

to denote the quark and gluon momenta respectively. The box graph contribution in Fig. 1 carries the same quantum numbers as the perturbative photon gluon fusion process, which has been discussed in the $g_1$ literature in connection with the contributions to $g_1$ from polarised glue and the axial anomaly [11-13,30,31]. We now calculate the contribution to $g_1$ from this process involving the exchange of two non-perturbative gluons, which provides a successful description of the physics of the non-perturbative pomeron. We do not attempt to interpret our calculation in terms of separate contributions from either the polarised glue or sea and the effect of the axial anomaly.

The non-perturbative gluon and quark propagators which appear in the vertical lines of Fig.1 are taken from the model of refs.[26,27]:

$$G_{\mu\nu}(l) = g_{\mu\nu} D(l^2) \qquad (7)$$

(where we work in Feynman gauge) and

$$Q(k) = (\gamma.k + m) S(k^2) \qquad (8)$$

(where $m$ is the constituent quark mass) respectively. Confinement is built into the model by requiring that $D(l^2)$ and $S(k^2)$ do not have any poles at time-like momentum on the physical sheet [26]. This requirement ensures that the moments of $D(l^2)$ and $S(k^2)$ are finite. The leading term in the hadronic cross section comes when the struck quark travels only a very short distance between interactions with the gluons [27]. This means that the horizontal quark lines in Fig.1 should be treated as perturbative; these quarks are placed on-shell when we take the imaginary part of the diagram.

The calculation of the two non-perturbative gluon exchange contribution to hadronic cross sections involves taking the first and second moments of $D^2(l^2)$ and $S^2(k^2)$. These moments are determined in terms of four parameters. The physics of the gluon propagator involves the coupling $\beta_0$ of the pomeron to a quark in the target and the mass parameter $\mu_0$ in the quark pomeron form-factor [28], which are determined from experiment to be $\beta_0 \approx 2.0$ GeV$^{-1}$ and $\mu_0 \approx 1$ GeV [29]. (These quantities are related to the values of the gluonic condensate $G_{\mu\nu}G^{\mu\nu}(0)$ in the vacuum and its correlation length [26].) The moments of the quark propagator $S(k^2)$ are determined by the vacuum quark condensate $<\text{vac}|\overline{q}q|\text{vac}> = -m_0^3$ where $m_0 \approx 225$ MeV and the constituent quark mass $m \approx 330$ MeV [27].

The spin-dependent part of the hadronic tensor in deep inelastic scattering enters as the anti-symmetric term in the proton line $\gamma_\rho(\gamma.p - \gamma.l)\gamma_\sigma$. This is

$$i\epsilon_{\rho\sigma\alpha\beta}(p-l)^\alpha \gamma_5 \gamma^\beta \to i\epsilon_{\rho\sigma\alpha\beta}(p-l)^\alpha s^\beta \qquad (9)$$

The anti-symmetric part of $W_{\mu\nu}$ is linear in the nucleon spin vector $s$. In order to pick out the part proportional to $g_1$ we choose to evaluate it with $s$ set equal to $p$ — even though this is unphysical it removes the contribution from the other structure function $g_2$ [14]. The leading term at $\nu \to \infty$ is independent of the nucleon mass, which we set to zero. The anti-symmetric part of the term that results from contracting equ.(9) with the trace of the quark loop is:

$$-4(m^2 + k^2)l^2(\xi - 2) \tag{10}$$

The delta function constraints on the horizontal lines evaluate as:

$$\int d^4l \ \delta((p-l)^2) = \frac{\pi}{2} \int d\xi \int dl^2 \tag{11}$$

and

$$\int d^4k \ \delta((k-l)^2)\delta((k+q)^2) = \frac{\pi}{2}\frac{1}{2\nu}\frac{1}{\xi} \int d(-k^2) \tag{12}$$

Inserting numerical factors, we find a contribution to the flavour singlet part of $g_1$:

$$-g^4\left(\frac{\pi}{2}\right)^2 2 \int_{x_{Bj}}^1 d\xi(1 - \frac{2}{\xi}) \int dl^2 l^2 D^2(l^2) \int_0^\infty d(-k^2)(m^2+k^2)S^2(k^2) \tag{13}$$

where $g$ is the quark-gluon coupling. The $l^2$ integral is [28]:

$$g^4 \int dl^2 l^2 D^2(l^2) = 18\pi \beta_0^2 \mu_0^2 \tag{14}$$

The integrand in the $k^2$ integral is the sum of two terms with opposite signs, which makes the value of this integral particularly model dependent. If we use the same model (non-perturbative) quark propagator

$$S(k^2) = \frac{4\pi^2}{3}\frac{m_0^3}{m^5} \exp \frac{k^2}{m^2} \tag{15}$$

as that used in [27], then the two non-perturbative gluon exchange process gives a contribution to the flavour singlet part of $g_1$, which is equal to

$$N\left(2\ln\frac{1}{x} - 1\right) \tag{16a}$$

with

$$N = \frac{\pi}{18}\beta_0^2\mu_0^2\left(\frac{m_0}{m}\right)^6 \simeq 0.09 \tag{16b}$$

The best fit of the form (16a) to the small $x$ SMC data yields $N = 0.085 \pm 0.01$, which corresponds to $\Sigma = 25 \pm 11\%$ [32].

The non-perturbative two gluon exchange process is a simple model of non-perturbative pomeron exchange in unpolarised hadronic interactions. In that case, it is necessary to correct the model result $x\overline{q} \sim 0.2$ by a factor $x^{-0.08}$ [27] to account for the fact that the pomeron is really more complicated than just two gluons, and reggeises. We do not know whether the two-gluon exchange contribution we have calculated for deep inelastic scattering also reggeises, or whether it is a fixed pole.

The Regge theory successfully describes the small $x$ part of the unpolarised structure function as measured by the NMC [15]. It seems reasonable that the same should be true of the polarised structure function in the same kinematic range of $x$ and $Q^2$. The gluonic exchange contribution to the flavour singlet part of $g_1$ in equ.(16) should be present in the SMC data and may be responsible for the rise in $g_1^p$ which is seen in the smallest $x$ bins. We show this in Fig. 2 where we plot our gluonic exchange contribution to $g_1^p$ together with the SMC data.

It is interesting to include the perturbative photon gluon fusion process into the model. Here the quarks carry large transverse momentum. We re-calculate the diagram in Fig. 1 using purely perturbative quark propagators and keep the quark transverse momentum $k_T^2 \geq \lambda^2$, where $\lambda^2 \geq 2$ GeV$^2$. In this case, equ.(13) is replaced by

$$\frac{1}{96\pi^2}g^2\alpha_s(\nu)\int dl^2 l^2 D^2(l^2) \int_0^1 d\xi\left(\ln\frac{2\xi\nu}{\lambda^2} - 1\right) \simeq 0.0008 \tag{17}$$

which makes a negligible contribution to $g_1$ at small $x$ compared to the non-perturbative contribution in equ.(16).

Finally, we discuss how our results may be manifest in the SMC and E-142 low $x$ data on the spin dependent nucleon structure function. The EMC and SMC data tells us that $g_1^p$ is positive in the $x$ range shown in Fig.2, whereas $g_1^n$ (which has been extracted modulo shadowing corrections) is negative over this range of $x$. This suggests that there is a large iso-triplet piece to $g_1$ at small $x$, which is described by equ.(5). The rise in $g_1^p$ which is seen in the smallest $x$ bins of the SMC may be due to the non-perturbative gluon exchange process which we have described here (see Fig. 2). This flavour singlet contribution will come to dominate the convergent iso-triplet contribution at very small $x$. Close and Roberts [32, 33] have also discussed a possible rising contribution to $g_1$ at small $x$ associated with the pomeron-pomeron cut [34]. At the present time the size of the experimental error and the theoretical uncertainty in the normalisation of equ.(16) mean that it is not possible to determine the relevant weights of the $a_1$ and gluon exchange contributions. This uncertainty means that the error which is quoted on tests of the Ellis-Jaffe sum-rule is probably too small (see also [32]). Of course, tests of the iso-triplet Bjorken sum-rule are independent of any theoretical discussion about the small $x$ behaviour of the flavour singlet part of $g_1$. It is sufficient within the present experimental error to use the same small $x$ extrapolation given in equ.(5) for both $g_1^p$ and $g_1^n$ in order to test the Bjorken sum-rule. At first sight, our result may appear disappointing in that it is harder than hitherto expected to make a precise test of the Ellis-Jaffe sum-rule. On the other hand, the real test for theoretical models is whether they can predict the structure function $g_1$ over a complete range of $x$ rather than just one moment! [11]

*Acknowledgements:*

We thank F. E. Close, N. N. Nikolaev, R. G. Roberts and A. W. Thomas for helpful discussions. This research is supported in part by the EU Programme "Human Capital and Mobility", Network "Physics at High Energy Colliders", contract CHRX-CT93-0537 (DG 12 COMA).

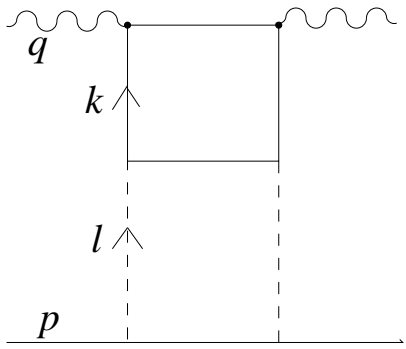

Fig. 1: The gluon exchange contribution to $g_1$.

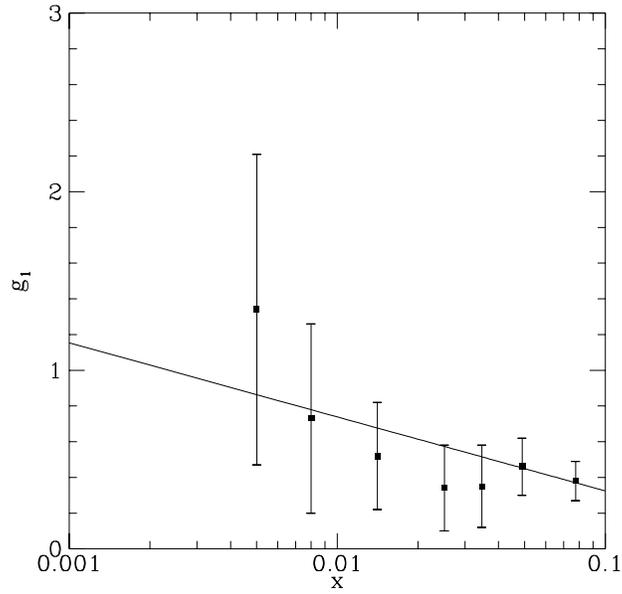

Fig. 2: The gluonic exchange contribution to $g_1^p$ at small $x$ (equ.(16)) together with the SMC data.